# The Result of the Neutrino-4 Experiment and the Cosmological Constraints on the Sterile Neutrino

A.P. Serebrov, R.M. Samoilov, M.E. Chaikovskii, O.M. Zherebtsov

*Petersburg Nuclear Physics Institute, National Research Center Kurchatov Institute, 188300, Orlova roscha 1, Gatchina, Russia*

E-mail: serebrov_ap@pnpi.nrcki.ru

## Abstract

We present a short discussion of the Neutrino-4 experimental results and the results of other experiments searching for the sterile neutrino. We estimated the contribution of the sterile neutrino with parameters $\Delta m_{14}^2 \approx 7.3 \, eV^2$ and $\sin^2 2\theta_{14} \approx 0.36$ obtained in the Neutrino-4 experiment to the energy density of the Universe. We address the contradiction between the measured sterile neutrino parameters and the constraints on the sterile neutrino from cosmology. With this article, we want to draw attention to the problem of the contradiction between experiment and theory, in order to inspire the search for theoretical models that include a sterile neutrino with mass in the region of several eV, and to the necessity to sufficiently increase the precision of the experiment.

Keywords: sterile neutrino, neutrino mass, neutrino mixing, dark matter.

## 1. Introduction

Starting from LSND in 2001 the significant amount of evidences of the sterile neutrino existence have piled up. Anomalies were observed in several accelerator and reactor experiments: LSND at a confidence level of 3.8 σ [1], MiniBooNE 4.7 σ [2], reactor anomaly (RAA) 3 σ [3,4], as well as in experiments with radioactive sources GALLEX/GNO, SAGE 3.2σ and BEST [5-7].

In 2018, in the Neutrino-4 experiment [8], a direct process of oscillations with a sufficiently high frequency was observed. This was the first direct observation of the effect of oscillations of reactor antineutrinos into a sterile state, in which it was possible to determine both the frequency and amplitude of the oscillations. The parameters of this process were measured to be $\Delta m_{14}^2 \approx 7.26 \text{ eV}^2$ and $\sin^2 2\theta_{14} \approx 0.38$ at CL $3.5\sigma$. A method of coherent summation of measurement results was proposed and implemented, which made it possible to directly observe the process of oscillations (see Fig. 1). In 2021, [9] a detailed description of the experiment was published, from preparatory operations to the final result, with a detailed analysis of possible systematic errors and MC simulations. The result was confirmed: at the level of $2.9\sigma$ with parameters $\Delta m_{14}^2 = (7.3 \pm 0.13_{st} + 1.16_{sys}) \text{ eV}^2$ and $\sin^2 2\theta_{14} = 0.36 \pm 0.12_{stat}$ $(2.9\sigma)$. Statistical analysis based on Monte Carlo gave a confidence level estimate of $2.7\sigma$.

The 2022 paper [10] presents an analysis of the results obtained in the Neutrino-4 experiment in comparison with the results of the NEOS, DANSS, STEREO, PROSPECT experiments on reactors, the MiniBooNE, LSND, MicroBooNE experiments on accelerators, with the IceCube experiment and the BEST experiment with $^{51}$Cr neutrino source. Combining the result of the "Neutrino-4" experiment and the result of the BEST experiment $\sin^2 2\theta_{14} = 0.34_{-0.09}^{+0.14}$, one can obtain more precise value of the oscillation amplitude $\sin^2 2\theta_{14} = 0.35_{-0.07}^{+0.09}$ $(4.9\sigma)$. It was shown that the results of the above-mentioned direct experiments on the search for sterile neutrinos are consistent within the framework of the 3+1 neutrino model within the available experimental accuracy. The sterile neutrino parameters make it possible to estimate the sterile neutrino mass $m_4 = (2.70 \pm 0.22)$ eV and the effective electron neutrino mass $m_{4\nu_e} = (0.82 \pm 0.16)$eV. Finally, one can present the PMNS matrix of the 3+1 neutrino model and the flavor mixing scheme.

It should be noted that the values of $\sin^2 2\theta_{14}$ obtained in reactor experiments conflict with the results of measurements of solar neutrinos interpreted on the basis of the Solar model. But the main problem for the obtained experimental result is the restriction on the sterile neutrino from cosmology. This article is devoted to review of this problem and determining the role of the sterile neutrino in cosmology.

## 2. Results of the Neitrino-4 experiment in comparison to other experiments

A detailed comparison of the results of the Neutrino-4 experiment with the results of other experiments is presented in our paper [10]. Here we restrict ourselves to only two illustrations: the demonstration of the experimental curve (fig.1) and the comparison of the Neutrino-4 result with the results of the other neutrino experiments (fig.2).

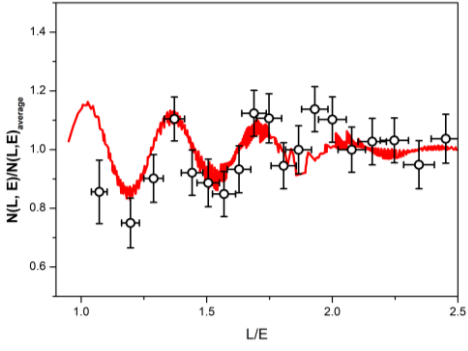

Fig 1 Neutrino signal oscillation curve. Red line – expected curve at $\Delta m_{14}^2 = 7.3 эВ^2$ and $\sin^2 2\theta = 0.36$, resolution 250 keV. $\chi^2/DoF = 20.61/17(1.21)$, GoF = 0.24 – for $m_{14}^2 = 7.3 eV^2, \sin^2 2\theta = 0.36$, $\chi^2/DoF = 31.90/19(1.68)$, GoF = 0.03 without oscillation.

In Fig. 2, the region of parameters of sterile neutrinos is highlighted in blue, and this region is determined by the experiments Troitsk, KATRIN, BEST, DANSS and The result of the experiment Neutrino-4 is in this region: $\Delta m_{14}^2 = (7.3 \pm 0.13_{st} + 1.16_{sys})\, eV^2$, $\sin^2 2\theta_{14} = 0.36 \pm 0.12_{stat}$ The red ellipse indicates 95% confidence level in the Neutrino-4 experiment, taking into account the systematic error. The result of the KATRIN experiment [11, 12] does not exclude the Neutrino-4 region.

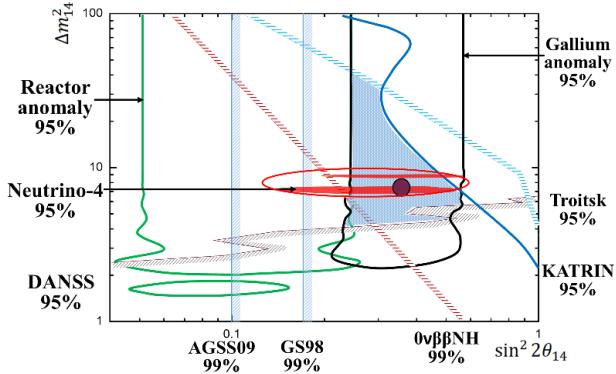

Fig 2 Comparison of the results of the Neutrino-4 experiment with the results of other experiments.

### 3. PMNS matrix in 3+1 neutrino model

In [10] the obtained oscillation parameters were applied to derive the PMNS matrix of the 3+1 neutrino model. We will use the matrix and flavor mixing scheme of the direct mass hierarchy for our analysis of the role of the sterile neutrino in cosmology (Fig. 3).

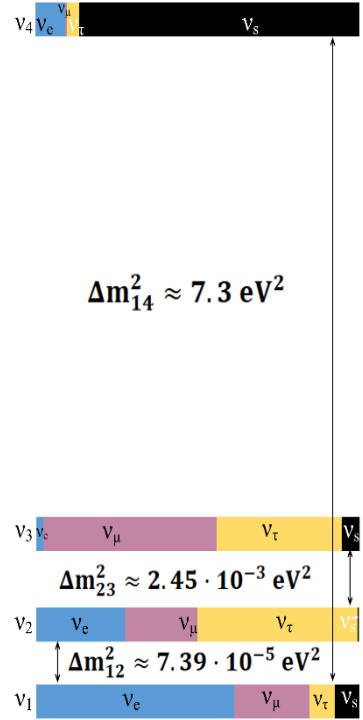

Fig 3 Scheme of mixing of active neutrinos and the sterile neutrino for normal mass hierarchy.

$$U_{PMNS}^{(3+1)} = \begin{pmatrix} 0.782^{+0.017}_{-0.016} & 0.524^{+0.017}_{-0.016} & 0.148^{+0.004}_{-0.004} & 0.301^{+0.035}_{-0.035} \\ 0.484^{+0.028}_{-0.034} & 0.473^{+0.027}_{-0.036} & 0.732^{+0.016}_{-0.025} & 0.074^{+0.021}_{-0.021} \\ 0.280 \div 0.330 & 0.678 \div 0.705 & 0.622 \div 0.657 & 0 \div 0.194 \\ 0.210 \div 0.273 & 0.060 \div 0.203 & 0.104 \div 0.236 & 0.931 \div 0.951 \end{pmatrix} \quad (1)$$

The most important features of this scheme for mixing the flavors of active neutrinos and sterile neutrino should be noted and they are illustrated in Fig. 3. First, the mass states $m_1, m_2, m_3$ are mixtures of electron, muon and tau flavors with a small fraction of the sterile state. Therefore, the $m_1, m_2, m_3$ mass states have a weak interaction, while the $m_4$ mass state is mostly sterile and has a weak interaction only due to the small contribution of electron, muon and tau flavors. As a result, the sterile neutrino $m_4$, which appears as a result of active neutrino oscillations, will propagate in the cosmic plasma for quite a long time before interaction and reverse transformation into an active neutrino. This circumstance creates the possibility of accumulation of sterile neutrinos and the possibility of their separation from the cosmic plasma at an earlier stage in comparison to active neutrinos.

### 4. Sterile neutrino role in cosmology

It is well known, that the process of neutrino oscillations in matter changes as a result of the interaction of neutrinos with matter (the MSW effect). This process is especially pronounced in space plasma. We begin our study of this process with the case of two neutrinos. Here and further we will rely on the well-known review publication [13] and the monograph [14]. In the two-neutrino case, the effective mixing matrix is determined by one angle

$$\begin{pmatrix} |\nu_e\rangle \\ |\nu_s\rangle \end{pmatrix} = \begin{pmatrix} \cos\theta_m & \sin\theta_m \\ -\sin\theta_m & \cos\theta_m \end{pmatrix} \begin{pmatrix} |\nu_1\rangle \\ |\nu_4\rangle \end{pmatrix} \quad (2)$$

The natural numbering of states is the one, in which in vacuum the effective states turn into the corresponding mass states: $|\nu_1\rangle \to |\nu_{m1}\rangle$ and $|\nu_4\rangle \to |\nu_{m4}\rangle$. In the mass basis, the neutrino Hamiltonian has the form:

$$H = H_m + U^+ V U \quad (3)$$

Where term $H_m$ contains energy differences based on the neutrino mass, and in the mass basis this matrix has form:

$$H_m = \text{diag}\left(\frac{m_1^2}{E}; \frac{m_4^2}{E}\right) \quad (4)$$

V – interaction matrix which has the diagonal form in the flavor basis and $U$ – mixing matrix.

For the two-neutrino case interaction matrix takes form:

$$V = \text{diag}\{V_e, 0\}$$
$$V_e = -25 \cdot 3.5 \cdot G_f^2 \cdot T^4 \cdot E \quad (5)$$

Here we consider only the second order part of the neutrino potential due to the small value of the baryon asymmetry. The MSW effect in the Sun occurs with the potential proportional to the first order of $G_f$ and has the form $V = \sqrt{2} G_f (n_e - n_{e^+})$ [13]. However, this potential appears in the assumption that there are no positrons in matter, which is true for the Sun, but in the early Universe the electron and positron densities are almost the same. Therefore, the first order term is cancelled out and one has to consider the second order term. In our calculations, we consider the value of the second order term (eq. 3) calculated in the work [14]. The main feature of the second order term is that it has the same negative sign for both neutrino and antineutrino. As a result, both particles have suppressed mixing angles at the early stages, where potential energy dominates the kinetic part. That means the MSW resonance do not occur in the early Universe, the mixing angle gradually increases with time, but never exceeds its vacuum value.

The plasma parameters gradually change, and therefore the effective neutrino levels transform accordingly. At each moment of time, we consider that the neutrino energy E is equal to $3.15T$, that is, the average energy for an ultra relativistic neutrino. The effective energy levels are the eigenvalues of the Hamiltonian and correspond to the eigenstates in the plasma at a given time. The difference between the levels, which determines the phase difference between propagating neutrino states in plasma, has the form:

$$E_2 - E_1 = \left(\left(\frac{\Delta m^2 \cos(2\theta)}{2E} - V\right)^2 + \left(\frac{\Delta m^2 \sin(2\theta)}{2E}\right)^2\right)^{1/2} \quad (6)$$

where $\Delta m_{14}^2 = 7.3$, $\sin^2 2\theta_{14} = 0.36$.

The level difference calculated using the presented expressions and parameters has a minimum near 0.0025 s.

Energy levels presented in Fig. 4 do not intersect, but come significantly closer. The minimal difference corresponds to the moment of time when the value of energy E in the equation (6) is small enough to get $\frac{\Delta m^2 \cos(2\theta)}{2E} > V$. In further development of the system the role of the potential becomes insignificant, the level is determined by the energy E, and the mixing angle approaches its vacuum value. It is at this moment that the interaction inside the plasma ceases to suppress the oscillation process.

The convenient characteristic of the mixing angle is $\sin^2 2\theta_m$, since the probability of oscillations at the times larger than oscillation periods is $\frac{1}{2}\sin^2 2\theta_m$. For this parameter, one can write a formula:

$$\sin^2 2\theta_m = \sin^2 2\theta \cdot \left(\left(\cos^2 2\theta - \frac{2 \cdot E \cdot V_e}{\Delta m^2}\right)^2 + \sin^2 2\theta\right)^{-1} \quad (7)$$

This expression monotonically depends on time (Fig.5):

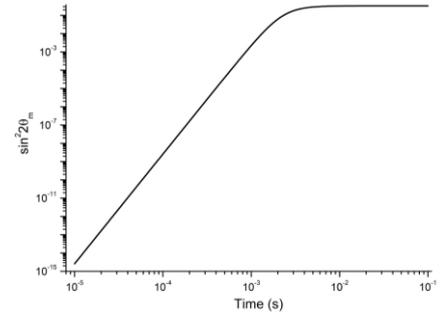

Fig 5 The behavior of the amplitude of electron neutrino oscillation into sterile state in cosmic plasma, depending on time.

In the 4-neutrino case we consider the following potentials from the works [14,15]:

$$\begin{aligned} V_e &= -3.5 \times 25 \times G_f^2 \times T^4 \times E \\ V_\mu &= -2 \times 25 \times G_f^2 \times T^4 \times E \\ V_\tau &= -25 \times G_f^2 \times T^4 \times E \\ V_s &= 0 \end{aligned} \quad (8)$$

It is considered that only relative potential energies influence neutrino mixing. Therefore, it is possible to make all the contributions positive by setting the potential with the smallest value $V_e$. equal to 0. Then the potentials will take a form:

$$\begin{aligned} V_e &= 0 \\ V_\mu &= 1.5 \times 25 \times G_f^2 \times T^4 \times E \end{aligned} \quad (9)$$

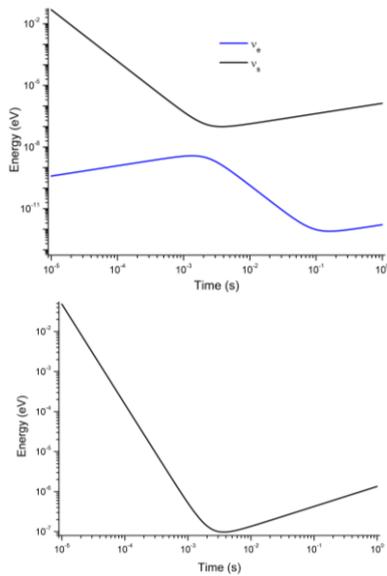

Fig 4 Top – Adiabatic levels for two-neutrino case, bottom Difference of the energy

$$V_\tau = 2.5 \times 25 \times G_f^2 \times T^4 \times E$$
$$V_s = 3.5 \times 25 \times G_f^2 \times T^4 \times E$$

We use that approximation in our calculations, but in fact, the muon potential on the temperature scale under consideration continuously changes. The multiplier is gradually converted from 1 to 3.5 with a decrease in the number of muons in the plasma.

In order to perform calculations, it is necessary to introduce vacuum mixing angles and neutrino masses. For the masses we take the values: $m_1 = 0.003$eV; $m_2 = 0.0091$eV; $m_3 = 0.0502$eV; $m_4 = 2.7$eV, so that $m_2^2 - m_1^2 = 7.38 \cdot 10^{-5}$eV$^2$ and $m_3^2 - m_2^2 = 2.44 \cdot 10^{-3}$eV$^2$.

Introducing the mixing matrix:

$$\begin{pmatrix} 0.784 & 0.525 & 0.1432 & 0.301 \\ -0.481 & 0.476 & 0.733 & 0.073 \\ 0.309 & -0.693 & 0.643 & 0.1 \\ -0.245 & -0.131 & -0.17 & 0.946 \end{pmatrix} \quad (10)$$

and considering its transformation due to the interactions in the cosmic plasma, we obtain the levels of neutrino energy showed in Fig. 6. Matrix elements values are presented without errors. The main contribution to the uncertainty of result is due to error of the $\sin^2 2\theta_{14}$.

The interaction of neutrino with cosmic plasma radically suppresses the process of oscillations, especially at the early stages. The effective mixing matrix gradually changes from the diagonal matrix at $t = 10^{-5}$ s:

$$\begin{pmatrix} 1. & 0. & 0. & 0. \\ 0. & 1. & 0. & 0. \\ 0. & 0. & 1. & 0. \\ 0. & 0. & 0. & 1. \end{pmatrix} \quad (11)$$

to the form which almost coincide with the vacuum mixing matrix at $t = 1\ s$:

$$\begin{pmatrix} 0.784 & 0.525 & 0.1432 & 0.301 \\ 0.481 & 0.476 & 0.733 & 0.073 \\ 0.309 & 0.693 & 0.643 & 0.1 \\ 0.245 & 0.131 & 0.17 & 0.946 \end{pmatrix} \quad (12)$$

Intermediate values are:

$$U_{ef}(10^{-4}s) = \begin{pmatrix} 1. & 0. & 0. & 0. \\ 0. & 1. & 0. & 0. \\ 0. & 0. & 1. & 0. \\ 0. & 0. & 0. & 1. \end{pmatrix} \quad (13)$$

$$U_{ef}(10^{-3}s) = \begin{pmatrix} 0.999 & 0.007 & 0.005 & 0.044 \\ 0.008 & 1. & 0.003 & 0.017 \\ 0.006 & 0.004 & 0.999 & 0.038 \\ 0.044 & 0.017 & 0.038 & 0.998 \end{pmatrix} \quad (14)$$

$$U_{ef}(10^{-2}s) = \begin{pmatrix} 0.953 & 0.048 & 0.024 & 0.299 \\ 0.075 & 0.988 & 0.11 & 0.073 \\ 0.05 & 0.121 & 0.986 & 0.1 \\ 0.29 & 0.079 & 0.12 & 0.946 \end{pmatrix} \quad (15)$$

$$U_{ef}(10^{-1}s) = \begin{pmatrix} 0.835 & 0.438 & 0.142 & 0.301 \\ 0.427 & 0.527 & 0.731 & 0.073 \\ 0.233 & 0.721 & 0.645 & 0.1 \\ 0.257 & 0.104 & 0.17 & 0.946 \end{pmatrix} \quad (16)$$

A fast change occurs in the region $10^{-3} - 10^{-1}$ s, in which the contributions of potential and kinetic terms become comparable.

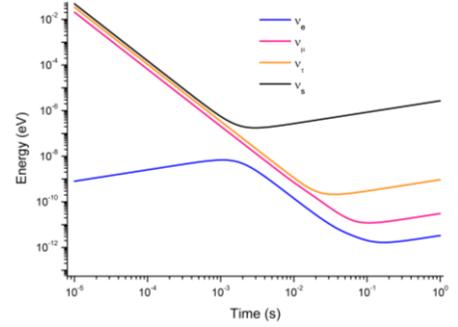

Fig 6 Behavior of adiabatic energy levels in the 4-neutrino system

We are interested in the density of sterile, tau, muon and electron neutrino at different points in time. In order to estimate these densities we should consider the dynamics of the processes of creation and destruction of various types of neutrino, solving the differential equation. We consider the behavior of neutrinos in the era after the annihilation of baryons and antibaryons (Fig.7). In this era, the contributions to neutrino interactions are made by the processes of scattering on electrons, positrons, neutrinos and antineutrinos, as well as the process of the annihilation of neutrino and antineutrino.

Time average probability of neutrino appearance due to oscillation is $\sim \frac{1}{2}\sin^2 2\theta$. Probability of neutrino generation is determined by collision frequency $1/\tau$ or mean free path which is proportional to the cross section and concentration $1/\tau = n_\nu \sigma$. Neglecting baryon asymmetry the total collision frequency for the electron neutrino can be expressed as [14,15]:

$$\frac{1}{\tau_{\nu_e}} = \Gamma_{\nu_e} = \frac{13}{9}\frac{7\pi}{24} G_f^2 T^4 E \quad (17)$$

For the $\nu_\mu$ and $\nu_\tau$ we use a smaller value of the collision frequency:

$$\frac{1}{\tau_{\nu_\mu}} = \frac{1}{\tau_{\nu_\tau}} = \Gamma_{\nu_\mu} = \frac{7\pi}{24} G_f^2 T^4 E \quad (18)$$

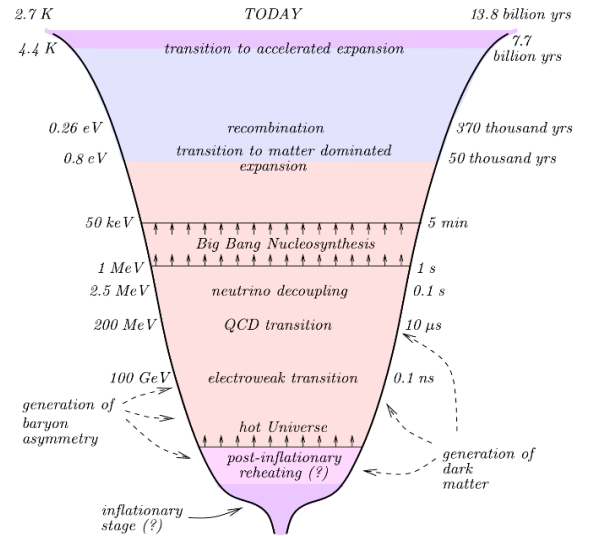

Fig 7 Stages of the Universe evolution from [14].

The dynamics of the sterile neutrino density is governed by three processes: 1) the expansion of the Universe, 2) transition of an active neutrino into the sterile state and 3) inverse transition of the sterile neutrino into an active state.

The transition of the sterile neutrino into an active one is considered as a two steps process: at first step the sterile neutrino oscillates into the active state with corresponding probability and then the active component take part in interaction.

Below we present the equation which taking into account $\nu_s$ generation and destruction. This equation (19) includes the effective interaction of sterile neutrino with plasma, due to oscillations:

$$\frac{dn_{\nu_s}}{dt} + 3Hn_{\nu_s} = \frac{1}{2}\left(\frac{\sin^2 2\theta_{m\,14}\, n_{\nu_e}}{\tau_{\nu_e}} + \frac{\sin^2 2\theta_{m\,24}\, n_{\nu_\mu}}{\tau_{\nu_\mu}} + \frac{\sin^2 2\theta_{m\,34}\, n_{\nu_\tau}}{\tau_{\nu_\tau}}\right) - \frac{1}{2}\left(\frac{\sin^2 2\theta_{m\,14}}{\tau_{\nu_e}} + \frac{\sin^2 2\theta_{m\,24}}{\tau_{\nu_\mu}} + \frac{\sin^2 2\theta_{m\,34}}{\tau_{\nu_\tau}}\right)n_{\nu_s} \quad (19)$$

where $n_{\nu_s}, n_{\nu_e}, n_{\nu_\mu}$ and $n_{\nu_\tau}$ are sterile, electron, muon and tau neutrino densities corresponding to the Fermi-Dirac distribution with zero chemical potential [14]. We used for the squares of the sine of the double angle the following values $\sin^2 2\theta_{14} = 0.36$, $\sin^2 2\theta_{24} = 0.024$ and $\sin^2 2\theta_{34} = 0.043$.

Equation (19) is a simplification that does not take into account the influence of oscillations of active neutrinos into sterile state on the density of active neutrinos themselves and considers only the oscillations to sterile neutrinos and oscillations of sterile neutrinos to active states. The initial condition is the zero sterile neutrino density at the starting time.

This equation is applicable up to the temperatures when the neutrino freeze-out, that is, to a temperature at which the density decreases so much that the interaction of neutrino with primordial plasma can be neglected. At this point, the interaction of neutrino with matter stops and only the expansion of space affects the further density dynamics. Value of $\sin^2 2\theta_{m\,14}$ is related to the vacuum value $\sin^2 2\theta_{0\,14}$ via equation (7) where due to absence of the interaction of the sterile neutrino with matter $V_e$ is equal to the interaction potential of the electron neutrino with plasma. Since $V_e$ is negative there is no MSW-resonance.

The rate of the sterile neutrino density increase is determined by the difference in the probability of appearance and disappearance of sterile neutrino. Both processes are proportional to the amplitude of electron neutrino oscillation into sterile neutrino (or vice versa) with a factor $1/2$. The process of appearance of the sterile neutrino is proportional to the density of electronic neutrinos $n_{\nu_e}$ and the interaction frequency of electronic neutrinos $\frac{1}{\tau_{\nu_e}}$. The process of transition of sterile neutrino into electronic neutrino is proportional to the density of sterile neutrinos $n_{\nu_s}$ and the interaction frequency of electron neutrinos $\frac{1}{\tau_{\nu_e}}$. Therefore, the factor $\frac{1}{2}\frac{\sin^2 2\theta_{m\,14}}{\tau_{\nu_e}}$ is included in both the generation and destruction of the sterile neutrino.

Electron neutrino density depends on temperature:

$$n_{\nu_e}(T) = \frac{3}{4}\frac{\zeta(3)}{\pi^2}T^3 \quad (20)$$

This equation has Hubble parameters H, which depends on the relativistic degrees of freedom. We use value 43/4 from PDG review for temperature less than muon mass. For the ultra-relativistic case the Hubble constant is related with temperature as following [14]:

$$H(T) = \frac{T^2}{M_{Pl}^*} \quad (21)$$

where $M_{Pl}^*$ is the reduced plank mass [14].

The magnitude of the plasma oscillation time is calculated by the formula

$$\tau_{osc} = \tau_0 \frac{\sin 2\theta_m}{\sin 2\theta_0} \quad (22)$$

Where $\tau_0$ — period of oscillations in vacuum $\tau_0 = \frac{4\pi E}{\Delta m^2}$, $\Delta m^2$ — The difference of squares of masses neutrino, $\sin 2\theta_m$ – sinus of the double mixing angle of two neutrinos in the plasma, $\sin 2\theta_0$ – sinus of the double mixing angle of two neutrinos in vacuum.

Fig 8 illustrates the relationship between oscillation frequency and collision frequency for neutrinos of different flavors as functions of time. The lower part of the figure 8 shows the dependence of the amplitude of oscillations between different flavors on time. We can note the presence of three critical moments when the oscillation amplitudes almost reach their level of the oscillation amplitude in vacuum and the time dependence drastically changes. These critical moments are also associated with the stabilization of the oscillation frequency at the level of the vacuum oscillation frequency, which increases due to a decrease of the neutrino energy due to the expansion of space. These are the moments at which neutrinos split from the plasma — the so-called moments of neutrino "freeze-out".

The time and temperature of freeze-out for different neutrinos can be estimated from the minima in the adiabatic levels time dependencies (see. Fig. 6) and from the minima of the oscillation frequency time dependencies in Fig. 8. It is at this moment the frequency and amplitude of the oscillations almost reach the level of vacuum parameters.

For a sterile neutrino, freeze-out occurs at $3 \cdot 10^{-3}\, s$, when the plasma temperature is $1.9 \cdot 10^{11}\, K$. For tau neutrinos, freeze-out occurs at $3 \cdot 10^{-2}\, s$, when the plasma temperature is $6 \cdot 10^{10}\, K$. For the muon neutrino, "freeze-out" occurs at $1 \cdot 10^{-1}\, s$, when the plasma temperature is $3.3 \cdot 10^{10}\, K$. For electron neutrinos, freeze-out occurs at $2 \cdot 10^{-1}\, s$, when the plasma temperature is $2.3 \cdot 10^{10}\, K$.

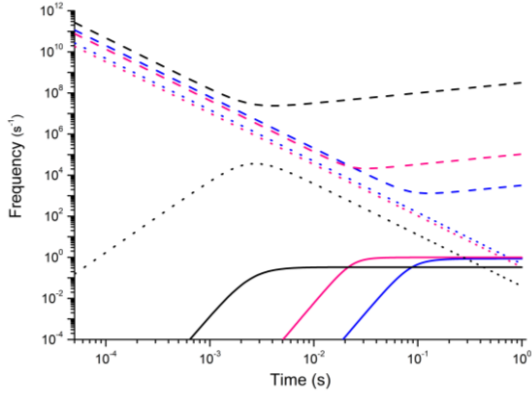

Fig 8 Neutrino "freeze-out" times. Short-dashed lines – collision frequency: blue – $\nu_e$, red – $\nu_\mu$, black – $\nu_s$ «collision» frequency. Dashed lines – oscillation frequency: blue –$\nu_e\nu_\mu$, red –$\nu_\mu\nu_\tau$, black –$\nu_e\nu_s$. Solid lines: blue – $\sin^2 2\theta_{21}$, red – $\sin^2 2\theta_{32}$, black – $\sin^2 2\theta_{14}$.

Fig 9 illustrates the rate of generation and destruction of sterile neutrinos with parameters obtained in the Neutrino-4 experiment. At some point in time the system reaches the balance between the processes of generation and destruction of the sterile neutrino.

As noted earlier, the mass states $m_1$, $m_2$, $m_3$ are a mixture of electron, muon and tau flavors with a small fraction of the sterile state. Therefore, the $m_1$, $m_2$, $m_3$ mass states have a weak interaction, while the $m_4$ mass state is mainly sterile and has a weak interaction only due to the small contribution of electron, muon and tau flavors. Ratio of the density of the sterile neutrinos to the density of electron neutrino at 1s is ~1. Ratio value is calculated with oscillation parameters from ref. [8].

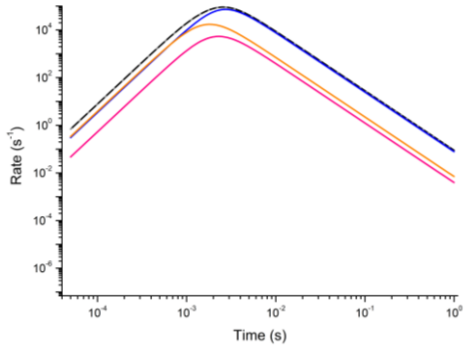

Fig 9 Sterile neutrino generation and disappearance. Black line – $\nu_s$ generation rate from $\nu_e, \nu_\mu, \nu_\tau$. Blue line – $\nu_s$ generation rate from $\nu_e$. Red line – $\nu_s$ generation rate from $\nu_\mu$. Yellow line – $\nu_s$ generation rate from $\nu_\tau$. Short-dashed line – total destruction rate $\nu_s$.

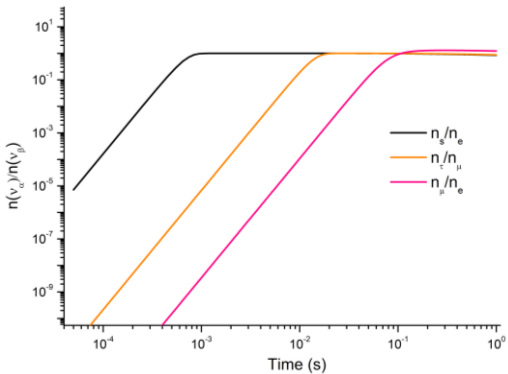

Fig 10 Neutrino relative densities.

The consequence of the considered processes is a very important result: by the time of freeze out of all neutrinos, the density of sterile, muon and tau neutrinos turns out to be almost the same as density of electron neutrinos. This result is illustrated in Fig. 10, which shows the dynamics of ratios of the densities for various neutrino pairs.

Now we should estimate the contribution of active and sterile neutrinos to the energy density of the Universe. It is quite obvious that the contribution of the sterile neutrino is dominant as their density is the same as densities of active neutrinos, but it has much higher mass $m_{\nu_4} = 2.7$eV. The contribution of active neutrinos to the energy density of the Universe is [14]: $\Omega_{\nu_1\nu_2\nu_3} \approx (m_{\nu_1\nu_2\nu_3}/1\text{eV}) \cdot 0.01h^{-2}$, where $h$ is the Hubble constant.

The contribution of the sterile neutrino to the energy density of the Universe is given by the equation:

$$\Omega_{\nu_4} \approx (\sum m_{\nu_i}/1eV)0.01h^{-2} \cdot n_{\nu_4} m_{\nu_4}/\sum(n_{\nu_i}m_{\nu_i})$$
$$n_{\nu_i} = n_{\nu_e}, \quad \sum(n_{\nu_i}m_{\nu_i}) = n_{\nu_e}\sum m_{\nu_i} \quad (23)$$
$$\Omega_{\nu_4} \approx (2.7\text{eV}/1\text{eV}) \cdot 0.01h^{-2} \cdot 1 \approx 0.053$$

and it equals to 5.3% of the total energy density of the Universe.

The presented calculation can be performed for other neutrino parameters. The fig 11 illustrates the dependence of the $\Omega_{\nu_4}$ on the sterile neutrino parameters obtained as a result of our analysis.

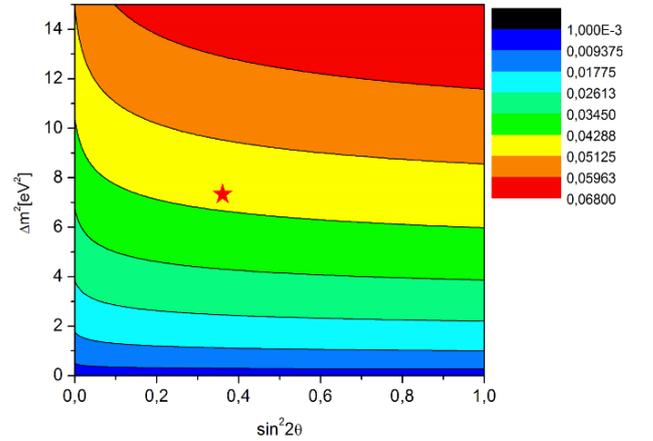

Fig. 11. Dependence $\Omega_{\nu_4}$ on $\Delta m_{14}^2$ and $\sin^2 2\theta_{14}$. Star is Neutrino-4 experiment best fit value

As a result of the above calculation, we derive an estimation of contribution of the sterile neutrino with parameters $\Delta m_{14}^2 = 7.3$ eV$^2$ and $\sin^2 2\theta_{14} = 0.36$ to the total energy density of the Universe. The obtained value is ~5% and it should be considered as a warm dark matter contribution to the present energy density. Also, within the discussed framework of the sterile neutrino density analysis we obtained the dependence of this contribution on $\Delta m_{14}^2$ and $\sin^2 2\theta_{14}$.

Prior to this, we were interested in estimating the contribution of a sterile neutrino with parameters close to those obtained in the Neutrino-4 experiment [8] to the dark matter. However, the above analysis can be extended to larger values of the sterile neutrino mass. We restrict ourselves to the version of equation (19) that includes only the electron and sterile neutrinos. It follows from the analysis of the equation that the contribution to dark matter in region $\sin^2 2\theta_{14} > 0.1$ depends mainly on the neutrino mass, and this fact illustrated in Figure 11. But if the mixing angle is small, its influence on the contribution to dark matter increases.

We are interested in establishing the mixing angles for heavy neutrinos which leave the contribution of the sterile neutrino to dark matter below the 25% limit. The calculation result is shown in Figure 12. This result demonstrates that heavy sterile neutrinos must have small mixing angles in order not to contradict the cosmological restrictions on the total contribution of dark matter to the energy density in the Universe. This form of the dependence of the angle on the mass can be simply explained.

Figure 10 shows that light sterile neutrinos at sufficiently large mixing angles come into thermodynamic equilibrium with plasma and their density becomes equal to that of active neutrinos. But for large masses, thermodynamic equilibrium is unacceptable, as it will lead to exceeding the threshold of 25%. This means that in order to remain within the bounds, as the neutrino mass increases, the mixing angle must decrease.

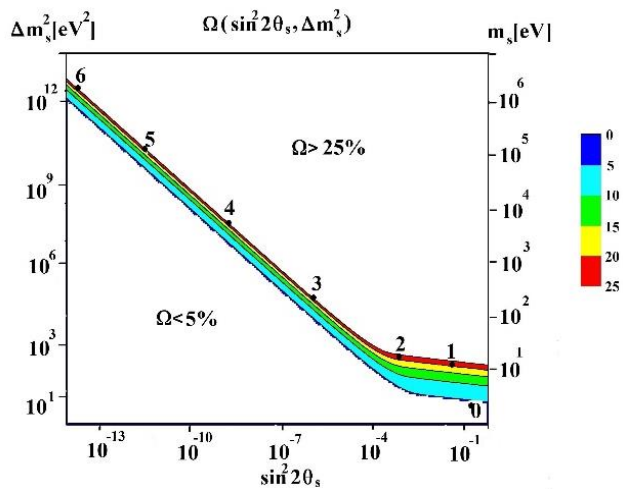

Fig.12. The area of parameters $\Delta m_{14}^2$ and $\sin^2 2\theta_{14}$ which lead to acceptable values of contribution to the dark matter.

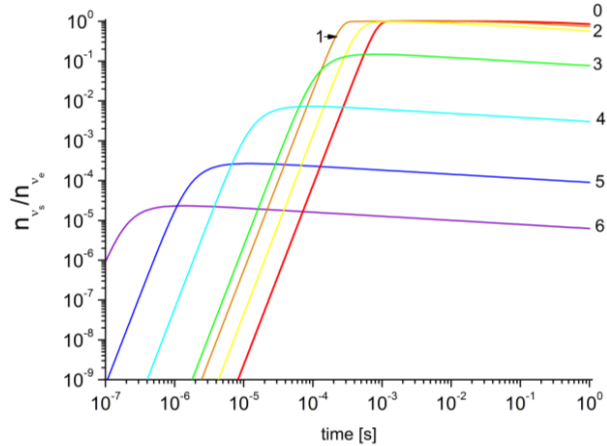

Fig.13. The ration of the number of sterile neutrinos to the number of electron neutrinos for several points in the $(\sin^2 2\theta_{14}, \Delta m_{14}^2)$ plane.

A smaller mixing angle results in the state where sterile neutrino does not have time to come into equilibrium with the electron neutrino before the moment of separation of the neutrino from the plasma. Therefore, the ratio $n_{\nu_s}/n_{\nu_e}$ remains less than 1. In figure 12, 7 points are selected on the plane $(\sin^2 2\theta_{14}, \Delta m_{14}^2)$, for which we plotted curves of the ratio of the number of sterile neutrinos to the number of electron neutrinos in figure 13. As the mixing angle decreases and the mass increases, this ratio decreases.

From the above analysis, we can conclude that heavy Dirac neutrinos should have a small mixing angle due to cosmological constraints. This means that heavy sterile neutrinos do not contribute to reactor neutrino experiments.

The following conclusions can also be drawn from the analysis.

1. A sterile neutrino with parameters $\Delta m_{14}^2 = 7.3 \; eV^2$, $\sin^2 2\theta_{14} = 0.36$ contributes approximately 5% to dark matter, but is relativistic and does not explain the structure of the Universe.

2. To explain the structure of the Universe, heavy sterile neutrinos with very small mixing angles are required.

3. Extension of the neutrino model by introducing two more heavy sterile neutrinos will make it possible to explain the structure of the Universe and bring the contribution of sterile neutrinos to the dark matter of the Universe to the level of 27%.

Above, we discussed the contribution of the sterile neutrino to dark matter and considered the constraints associated with the total energy of dark matter. We came to the conclusion that the parameters obtained in the experiment do not contradict the constraints on the energy density, and moreover, there is room for the introduction of heavier sterile states.

However, there are other constraints based on cosmological models and observations. There are three types of observations that impose limits on sterile neutrino parameters: 1) primordial nucleosynthesis and distribution of light nuclei [16, 17] 2) cosmic microwave background (CMB) [18] 3) clustering of large-scale cosmological structures [19, 20].

The introduction of a sterile neutrino with a mass of the order of eV to the model of the development of the early Universe changes the number of relativistic degrees of freedom during the period of nucleosynthesis and affects the expansion rate of the Universe at the moment of separation of photons from matter. As a result, a sterile neutrino shifts the moment of neutron freeze-out, and therefore affects the ratios of light nuclei in the universe. The anisotropy of the microwave background also turns out to be sensitive to the sterile neutrino parameters. The effect of neutrinos is usually expressed in terms of the effective number of relativistic degrees of freedom $N_{eff}$. The model with three active neutrinos predicts $N_{eff}^{3\nu} = 3.046$. The recently calculated constraints on the effective number of degrees of freedom obtained from fitting data on light elements in the universe are $N_{eff} = 2.843 \pm 0.154$ [21], and observations of the microwave background lead to the value $N_{eff} = 2.99 \pm 0.17$ [18]. These results are in good agreement with the 3 active neutrino model and leave open only a limited range of parameters for sterile neutrinos.

In models with a heavy unstable neutrino that decays as a result of mixing with active neutrinos, restrictions on the masses and mixing angles for heavy neutrinos appear. The decay of a heavy neutrino into a light neutrino and a gamma quantum gives rise to radiation of a certain energy equal to $m_s/2$. The mixing angle in this case determines the intensity of the decay, and the concentration of such neutrinos at the moment of neutrino freeze out, and therefore ultimately determines the intensity of the emitted gamma quanta. Observations of the gamma-ray spectrum in the range of several tens of keV impose constraints on decaying neutrinos [22]. These limitations are also shown in Figure 14. The result of experimental observations, the so-called 3.5 keV anomaly, which was closed by subsequent experiments [23], is also shown here.

There is a method to search for the sterile neutrinos with energies of the order of several keV in experimental facilities. The existence of such a sterile neutrino distorts the β-decay spectrum, and therefore can manifest itself in experiments on the direct measurement of the electron neutrino mass, based on a detailed study of the β-spectrum of tritium decay. At the moment, the best result in experiments of this type was obtained by the KATRIN collaboration [12].

The experimental constraints on the eV and keV sterile neutrinos are considered in KATRIN collaboration publication [24]. In figure 14 we illustrate the already excluded regions in parameter space and the sensitivity region of the KATRIN experiment. In the eV region the KATRIN experiment has perspective to either confirm or disprove our result, however, in the keV region the sensitivity of the KATRIN experiment is yet insufficient to reach the area where the sterile neutrino could be considered as candidate to Dark Matter particle.

Current restrictions on the parameters of keV sterile neutrinos obtained in experiments on measuring the mass of neutrinos are rather weak, and cosmological restrictions turn out to be much stronger (Fig.14). The range of parameters that the KATRIN collaboration plans to explore reaches mixing angles up to $\sin^2 2\theta \sim 10^{-8}$ [24]. This region is indicated in Figure 14. Although we believe that direct laboratory experiments are more preferable than cosmological observations and models, we should note that the sensitivity region of the KATRIN experiment for keV neutrinos is excluded by constraints on the total density of sterile neutrinos in the universe.

Nevertheless, we believe that these studies should be continued. As in the case of a sterile neutrino in the range of parameters of the Neutrino-4 experiment, direct observation of a signal prohibited by cosmology in a laboratory experiment will require revising cosmological models and their theoretical premises.

The sterile neutrino with mass of several eV also distorts the β-decay spectrum. In this region, the limitations obtained so far in the KATRIN experiment [24] (Fig. 14) do not contradict the results of the Neutrino-4 experiment. The measurement method used in the KATRIN experiment has a maximum sensitivity in the region of $100-1000$ eV$^2$, and in the region of several eV, reactor experiments turn out to be more efficient.

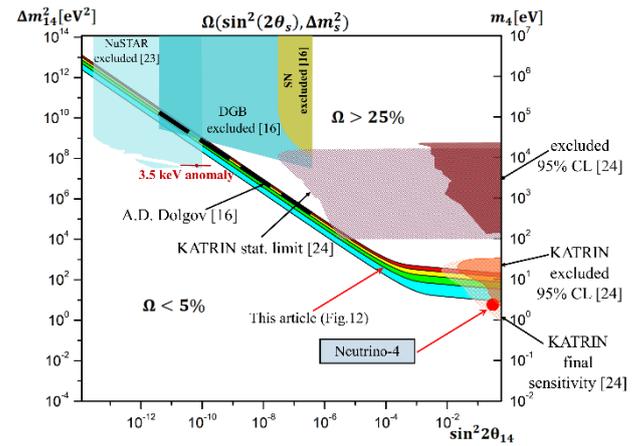

Fig 14. Constraints on the sterile neutrino parameters.
1) Red spot – result of the Neutrino-4 experiment; 2) This article (Fig.12) – area of the $\Omega_s$ values in 5-25% range; 3) A.D. Dolgov [16] – result from the Ref.[16] for $\nu_e - \nu_s$ mixing with $\Omega_s = 30\%$ (it should be noted that result of calculation presented in this work based on equation (19) is consistent with A.D. Dolgov results presented in Ref.[16]; 4) DGB – experimental constraints based on diffuse gamma background; 5) experimental constraints from SN1987 observation; 6) constraints from NuSTAR experiment [23]; 7) KATRIN excluded 95% CL – constraints on eV-scale sterile neutrino from KATRIN experiment; 8) KATRIN final sensitivity – sensitivity limit of the KATRIN experiment for eV-scale sterile neutrino; 9) excluded 95% CL – constraints from neutrino mass measurements experiment from Ref. [24]; 10) KATRIN stat. limit[24] – statistical limit of the KATRIN experiment for keV-scale sterile neutrino

The cosmological constraints conflict with the experimental data obtained at the Neutrino-4 experiment. But it should be noted that, any conclusions from cosmological observations are model dependent. Calculations of primordial nucleosynthesis and microwave background require a large number of parameters that cannot be measured directly, and are carried out taking into account some hypotheses regarding the composition of the primordial plasma. For example, it is assumed that the magnitude of the lepton asymmetry is negligible - $10^{-9}$, and chemical potentials of particles can be considered to be zero. If we consider the system without these conditions, then the neutrino potential must take into account the contribution of the first order in the constant $G_f$. The first-order contribution has the form [16]: $V_f = 0.95 \times G_f \eta T^3$, where $\eta$ is the magnitude of the charge asymmetry. For an electron neutrino $\eta = 2\eta_{\nu_e} + \eta_{\nu_\mu} + \eta_{\nu_\tau} + \eta_e - \eta_n/2$, and individual asymmetries for each type of particles are defined as the ratio of the difference in the number densities of particles and antiparticles to the number density of photons: $\eta_x = (n_x - n_{\bar{x}})/n_\gamma$. This additional contribution does not depend on the neutrino energy and depends on the temperature as $T^3$. Together with a small value of asymmetry, at high temperatures the contribution of the second order in $G_f$ turns out to be dominant, therefore, in calculations within the standard cosmology, the contribution of the first order can be neglected.

If we consider a sufficiently large asymmetry, then the diabatic energy levels of active and sterile neutrinos can intersect, which will lead to resonant oscillations into a sterile state, by analogy with resonant oscillations between electron and muon neutrinos in the Sun (MSW resonance).

We considered the potentials:
$V_e = 0{,}95 \times G_f \eta T^3 - 3.5 \times 25 \times G_f^2 \times T^4 \times E$
$V_s = 0$

for different values of $\eta$ and obtained corresponding curves for ratio of sterile and active neutrino densities as function of time. We considered a few values from $10^{-9}$ to $10^{-7}$s. With the standard cosmological value $\eta = 10^{-9}$ the first order term does not affect the dynamics of the neutrino densities, which is consistent with the hypothesis that this contribution can be neglected. It turned out that even with an increase in the asymmetry up to $\eta = 10^{-7}$ the contribution of the first order in $G_f$ term does not affect the thermalization dynamics.

However, for the values $\eta = 10^{-1}$ and $\eta = 1$ ratio of the sterile neutrino to electron neutrinos is 0.1 and 0.01 respectively in the time interval 1 – 100 s (Fig. 15).

Therefore, in these cases, sterile neutrinos have small influence on nucleosynthesis, and the contribution of sterile neutrinos to dark matter will be 0.5% and 0.05%, respectively. Thus, a sterile neutrino with parameters $m_{14}^2 \approx 7.3 \, eV^2$ and $\sin^2 2\theta_{14} \approx 0.36$ does not contradict the observed nucleosynthesis, if such large values of lepton asymmetry can be substantiated. In this regard, it can be noted that in the work of A. D. Dolgov "Neutrino oscillations in the early Universe. Resonance case" [25] similar situation was considered. The idea of that work is the transition $\nu_\alpha \to \nu_s$ may be more favorable than the transition of the corresponding antineutrinos. The feedback is positive and leads to a further increase in asymmetry and makes the transition $\nu_\alpha \to \nu_s$ more and more efficient compared to $\bar\nu_\alpha \to \bar\nu_s$. The lepton asymmetry generated in the early Universe by neutrino oscillations on sterile partners reaches the asymptotic values of the asymmetry at the 0.2 – 0.3 [25]. Of course, a detailed consideration of such a scenario with experimental oscillation parameters is required.

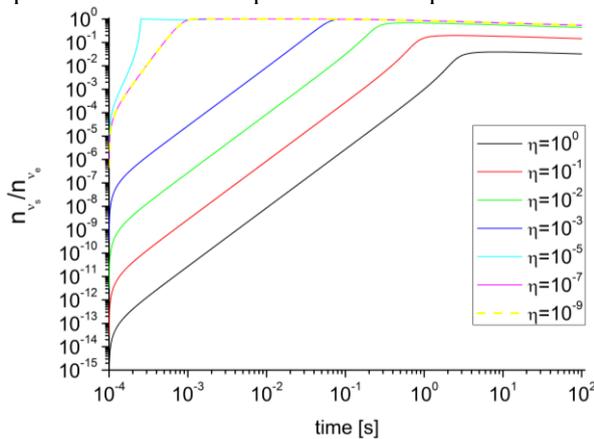

Fig.15. The ratio of sterile neutrino density to the electron neutrino density for the several values of the parameter η.

Another option for expanding the theory is to take into account the possibility of the decay of a light sterile neutrino. In [23], a formula is given for the decay rate of a sterile neutrino into an active neutrino and a photon:

$$\Gamma_s = \frac{9}{1024}\frac{\alpha}{\pi^4} G_F^2 m_s^5 \sin^2 2\theta$$
$$= 1.38 \times 10^{-22} \left(\frac{m_s}{1 keV}\right)^5 \sin^2 2\theta \; s^{-1}$$

This is a very small value, and decay with such parameter will not make any noticeable changes in the above calculations. Therefore, we propose to consider the hypothetical possibility of the fast decay of a sterile neutrino without discussing the mechanisms and products of this decay, since we are only interested in the effect of the decay rate on the number of sterile neutrinos in the early universe.

A natural limitation from the experiment is the fact that the neutrino leaves an oscillatory signal in our experimental setup, so it does not decay over a length of about 10 meters. From this we get the lower limit of the decay period of the order $\tau_0 = 2 \times 10^{-14} \, s$. If in addition we consider that reactor anomaly can be observed at distances up to 1 km, then the lower limit can be increased to $\tau_0 = 2 \times 10^{-12} \, s$. Finally, an estimate can be made if we assume that the effect of oscillations is also observed in the IceCube [26] experiment, although the errors are still quite large ($\Delta m_{24} = 6.7^{+3.9}_{-2.5}$ eV$^2$, $\sin^2 2\theta_{24} = 0.33^{+0.33}_{-0.17}$), then considering the distance (diameter of the Earth) and neutrino energy (~100 GeV) we get $\tau_0 = 2 \times 10^{-12} \, s$.

Including this decay into equation (20) as an additional channel for sterile neutrino losses, we get the value $n_s/n_e \ll 0.01$ by the start of the nucleosynthesis at $t = 1 \, s$. Density ratios of the sterile neutrino to the electron neutrino for various decay periods is shown in figure 16. The calculation result shows that in a wide range of values of $\tau_0$ it is possible to get the contribution of the sterile neutrino to dark matter at a level that does not contradict cosmological constraints. For example, $\tau_0 = 10^{-7} \, s$ leads to $n_s/n_e \approx 0.1$. The value $\tau_0 = 10^{-7} \, s$ can be considered as an upper limit on the decay period of a sterile neutrino, derived from cosmological constraints based on primordial nucleosynthesis. With this simple argument, we want to show that there are ways to consistently include the sterile neutrino with the parameters obtained in the experiment in cosmology, but for this it will be necessary to significantly expand the theoretical model.

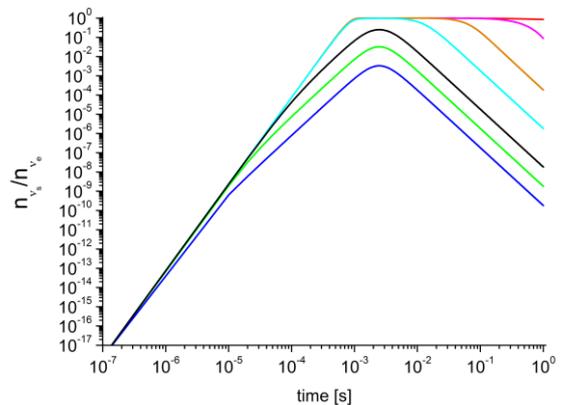

Fig. 16. The ratio of sterile neutrino density to the electron neutrino density, taking into account the decay of a sterile neutrino. The lifetime of a sterile neutrino in the comoving coordinate system is $\tau_0$. Oscillation parameters are $\sin^2 2\theta_{14} = 0.36$, $\Delta m_{14}^2 \approx 7$ eV$^2$.

If the Neutrino-4 result is confirmed at a confidence level of more than 5σ at our new experimental facility with three times the sensitivity, as well as by other scientific groups, then the above theoretical limitations will have to be revised. Since direct observation of a sterile neutrino in a laboratory experiment can become the defining criterion, a significant revision of the entire model of the dynamics of the early universe will be required.

With this article, we want to draw attention to the problem of the contradiction between experiment and theory, in order to inspire the search for theoretical models that include a sterile neutrino in the region $\Delta m_{14}^2 \sim$ 5-10 eV$^2$ and $\sin^2 2\theta_{14} \sim$ 0.3-0.4. In particular, we would like to note that the splitting of such a sterile neutrino from plasma occurs earlier by approximately two orders of magnitude with respect to active neutrinos. As can be seen from fig. 6 and 8, the moment of splitting of the sterile neutrino from the plasma is approximately 2 10$^{-3}$ s, and for the active electron neutrino 2 10$^{-1}$ s. By the moment of freezing the ratio of neutrons and protons and the beginning of primordial nucleosynthesis, the sterile neutrino practically no longer interacts with plasma and has little effect on that process. It remains unclear how the above constraints change when this fact is taken into account.

Let's consider the calculation of the $^4$He mass content using the well-known calculation scheme from [14]

$$^{4\text{He}}Y = m_{^4\text{He}} \cdot \frac{n_{^4\text{He}}(T_{ns})}{m_p \left(n_p(T_{ns}) + n_n(T_{ns})\right)} = \frac{2}{n_p(T_{ns})/n_n(T_{ns}) + 1}$$

$$n_n(T_{ns})/n_p(T_{ns}) = n_n/n_p \cdot e^{-t_{ns}/\tau_n} = e^{-(m_n - m_p)/T_n} \cdot e^{-t_{ns}/\tau_n}$$

The mass content of $^4$He is determined by the ratio of neutrons to protons at the moment of neutron freeze-out at a cosmological time of approximately 1.2 s and the nucleosynthesis time of approximately 4 minutes.

The number of degrees of freedom at the moment of neutron freeze-out is equal to:

$$g_*^{T_n} = 2 + \frac{7}{8} \cdot 4 + \frac{7}{8} \cdot 2 \cdot N_\nu$$

The first contribution arises from photons, the second from electrons and positrons, and the third is due to light neutrinos. Accordingly, for $N_\nu = 3, g_*^{T_n} = 10.75$, and for $N_\nu = 4$, $g_*^{T_n} = 12.5$. Although the number of degrees of freedom increases by 16.3%, the expansion rate of the Universe increases by 7.8%, because square root dependency. For the radiation-dominated era the Friedman equation has the form: $H^2 = \frac{8\pi^3}{90} G g_* T^4$, where the space expansion is determined by the Hubble parameter, and the plasma cooling rate is determined by the number of relativistic degrees of freedom. Thus, the rate of expansion of the Universe during the period of neutron hardening increases by 7.8% due to the introduction of an additional light neutrino.

The number of degrees of freedom at the time of nucleosynthesis is:

$$g_*^{T_n} = 2 + \frac{7}{8} \cdot 2 \cdot N_\nu \cdot \left(\frac{4}{13}\right)^{4/3}$$

because only two types of relativistic particles affect the expansion rate of the Universe: photons and neutrinos, and, since neutrinos no longer interact with plasma, their effective contribution is suppressed [14] and then for $N_\nu = 3, g_*^{T_n} = 3.36$, and for $N_\nu = 4, g_*^{T_n} = 3.81$. Although the number of degrees of freedom increases by 13.5%, the expansion rate of the Universe increases by 6.5%, because square root dependency. Thus, the rate of expansion of the Universe during the period of nucleosynthesis increases by 6.5% due to the introduction of an additional light neutrino. The average value of the increase factor for the interval from 1.2 s to 265 s is approximately 7%.

Finally, already in the recombination era (T = 0.26 eV) and later, a light sterile neutrino with a mass of 2.7 eV becomes nonrelativistic and does not contribute to the number of degrees of freedom. Therefore, the number of degrees of freedom at the moment of recombination is:

$$g_*^{\text{rec}} = 3.36$$

and remains the same for all subsequent stages of the development of the Universe. This is a very important point to mention.

Let's note that the Hubble constant and the number of degrees of freedom during the period of neutron freeze-out and nucleosynthesis are related by a simple relation $H_0/\sqrt{g_*}$, because the plasma temperature is determined by the space expansion rate and the plasma cooling rate, which depends on the number of relativistic degrees of freedom. It is important to clarify that the relationship between these quantities for the epoch of radiative dominance is given by the Friedmann equation: $H^2 = \frac{8\pi^3}{90} G g_* T^4$ and hence the strict relation $H_0/\sqrt{g_*}$. The value of the Hubble constant $H_0$ = 67.4 ± 0.5 (km/s)/Mpc [18] presented by the Planck collaboration refers to the early Universe and is based on $N_{\text{eff}}$ = 2.96 ± 0.15. If we assume that the processes of neutron hardening and nucleosynthesis occurred in the presence of the fourth neutrino, then it is necessary to introduce the justified earlier correction of 7% for the value $H_0$ = 67.4 ± 0.5 (km/s)/Mpc, then, $H_0^*$ = 72.1 ± 0.9(km/s)/Mpc. Thus, the value of the Planck collaboration ($H_0$ = 67.4 ± 0.5 (km/s)/Mpc corrected for the fourth neutrino is $H_0^*$ = 72.1 ± 0.9(km/s)/Mpc are consistent within one standard deviation with the result of the Riess group ($H_0$ = 73.2 ± 1.3 (km/s)/Mpc) [27].

Thus, using Fig. 10 from the [28], we can add to this picture the probability density of the value $H_0 (T_n \div T_{ns})$ = 72.1 ± 0.9 (km/s)/Mpc for the interval of cosmological times of neutron freeze-out and nucleosynthesis considering sterile neutrino.

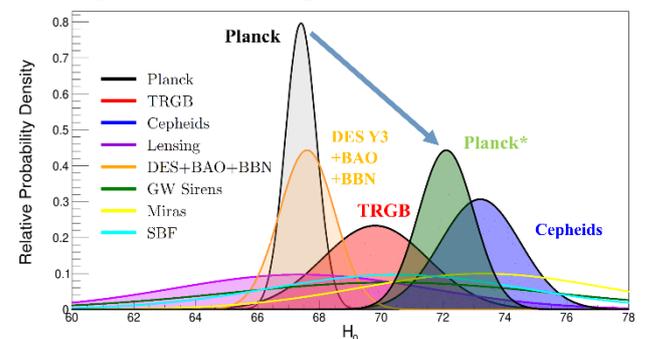

Fig. 17. The introduction of a correction for the fourth neutrino makes it possible to reconcile the value $H_0$ = 67.4 ± 0.5 (km/s)/Mpc from the Planck collaboration with the value $H_0$ = 73.2± 1.3 (km/s)/Mpc from the Riss group. Data taken from [28].

This is a very important conclusion. The introduction of the fourth neutrino removes the Hubble "tension" problem. The importance of the conclusion is not just that the Hubble "tension" problem is removed, but that this required a fourth neutrino.

The thing is that the Planck collaboration deliberately relies on the model of three neutrinos and measurements of the mass content of $He^4$ by Aver group [29] and neglects the results of other measurements, for example by Izotov group [30].

**Acknowledgements**

The work was supported by the Russian Science Foundation under Contract No. 20-12-00079. The authors are grateful to V.A. Rubakov, A.D. Dolgov and Z.G. Berezhiani for advice and comments on the theoretical aspects of this work. The authors are grateful to the colleagues of PNPI NRC KI and INR RAS for useful discussions at the seminars